\documentstyle[seceq,epsf,mbf]{ptptex}

\title{Effects of neutrino oscillation on supernova neutrino:
inverted mass hierarchy}

\author{
K. {\sc Takahashi}$^*$ and K. {\sc Sato}$^{*,**}$}

\inst{
$^*$Department of Physics, University of Tokyo,
\\
7-3-1 Hongo, Bukyo, Tokyo 113-0033, Japan
\\
$^{**}$Research Center for the Early Universe, University of Tokyo, 
\\
7-3-1 Hongo, Bukyo, Tokyo 113-0033, Japan
}

\recdate{
}

\abst{
We study the effects of neutrino oscillation on supernova neutrino
in the case of the inverted mass hierarchy ($m_{3} \ll m_{1} < m_{2}$) 
as well as the normal mass hierarchy ($m_{1} < m_{2} \ll m_{3}$). 
Numerical analysis using a realistic supernova and presupernova model allows 
us to discuss quantitatively a possibility to probe neutrino oscillation 
parameters. We show that information about the mass hierarchy can be obtained 
if $\theta_{13}$ is rather large ($\sin^{2}{2 \theta_{13}} > 10^{-3}$) and 
that $\theta_{13}$ can be well probed by SuperKamiokande if the neutrino 
mass hierarchy is the inverted case. Errors due to the uncertainty of the 
original neutrino spectra and the Earth effect are also discussed.
}

\begin{document}

\maketitle

\section{Introduction}

Neutrino mixing and mass spectrum are the keys to probe new physics
beyond the standard model of particle physics. Some of the neutrino
oscillation parameters have been revealed dramatically by the observation
of the atmospheric neutrino \cite{Fukuda1999} and the solar neutrino
\cite{Fukuda2001,SNO,Bahcall2002,Barger2002Sun,Holanda2002}. 
Recently the first results of the KamLAND experiment have confirmed
the Large Mixing Angle (LMA) solution of the solar neutrino problem
\cite{KamLAND}. An upper bound on $\theta_{13}$ has also been
obtained from CHOOZ experiment \cite{Apollonio1999} and a lower bound is
expected to obtained from single and double beta decay experiments
\cite{MinakataSugiyama}. But there still remain some ambiguities 
in the properties of neutrinos: the mass hierarchy, i.e., normal or
inverted and the magnitude
of $\theta_{13}$.  Current status is reviewed by many authors 
\cite{Raffelt2002,Cei2002,Raffelt2003,Bilenky2002}.

In such present situation, much attention have been paid to another 
neutrino source, supernova. This is a completely different system from sun, 
atmosphere, accelerator, and reactor in regard to neutrino energy, 
flavor of produced neutrinos, propagation length and so forth. Then 
neutrino emission from a supernova is expected to give valuable information 
that can not be obtained from neutrinos from other sources.
In fact, pioneering observations of neutrinos from SN1987A
\cite{K2_SN1987a,IMB_SN1987a} contributed significantly to our knowledge of 
the fundamental properties of neutrinos \cite{Arafune,Sato,Goldman}. 
Especially there have been many studies about the implication for the 
mass hierarchy from the observed neutrino events and the inverted
hierarchy is disfavored if $\sin^{2}{\theta_{13}}$ is rather large
($> 10^{-4}$) \cite{Jegerlehner1996,LunardiniSmirnov,Minakata2001}.
Here, normal and inverted mass hierarchies are the mass pattern
$m_{3} \gg m_{2} > m_{1}$ and $m_{2} > m_{1} \gg m_{3}$, respectively.
In our notation $\Delta m^{2}_{ij} = m^{2}_{j} - m^{2}_{i}$, 
$\Delta m^{2}_{12}$ and $\Delta m^{2}_{13} \sim \Delta m^{2}_{23}$ are
the mass squared differences which are related with the solutions of the 
solar and atmospheric neutrino problems, respectively. There have also been 
studies to try to extract the unknown neutrino properties from future 
supernova 
\cite{Dighe2000,Fogli2002,LunardiniSmirnov2003,DigheKeilRaffelt1,DigheKeilRaffelt2,KT}.

In this paper, we calculate numerically the effects of neutrino oscillation
on supernova neutrino, extending our previous study \cite{KT} where all 
analyses are performed with normal mass hierarchy, and
investigate the possibility to identify the mass hierarchy and to probe 
the neutrino oscillation parameters $\theta_{13}$ by the 
observation of the neutrinos from the next galactic supernova. 
We use the original neutrino spectra from supernova based on a realistic 
supernova model and the density profile of the progenitor star based on a 
realistic presupernova model. Since uncertainties of the original neutrino
spectra are important in this analysis, we estimate the effect of them 
on our analysis. The Earth matter effect is also discussed, which have 
already been studied in the case of normal hierarchy in our previous
papers \cite{KTearth,KT3}. 

This paper is organized as follows. In section II we summarize the
properties of supernova neutrino briefly. The method of
analysis is described and the results are shown in section III.
We discuss some ambiguities in the basis of our study and summarize
our results in section IV.

\section{Supernova Neutrino}

Here we summarize the properties of supernova neutrino.
For details, see, for example, a review by Suzuki\cite{Suzuki}.
Almost all of the binding energy of the neutron star,
\begin{equation}
E_{\rm b} \simeq \frac{G M_{\rm NS}^{2}}{R_{\rm NS}}
\simeq 3 \times 10^{53} {\rm erg} 
\left( \frac{M_{\rm NS}}{M_{\odot}} \right)^{2}
\left( \frac{10 {\rm km}}{R_{\rm NS}} \right),
\end{equation}
is radiated away as neutrinos. Here $G$, $M_{\rm NS}$ and $R_{\rm NS}$ are
the gravitational constant, the mass and radius of the neutron star, 
respectively. Due to the difference of interaction strength, average 
energies are different between flavors. Although quantitative estimate of 
the difference is difficult, it is qualitatively true that 
$\langle E_{\nu_{e}} \rangle <  \langle E_{\bar{\nu}_{e}} \rangle 
< \langle E_{\nu_{x}} \rangle$.
Here $\nu_{x}$ means $\nu_{\mu},\nu_{\tau}$ and their antineutrinos.
These differences are essential in this paper.

\begin{figure}[t]
\epsfxsize=12cm
\centerline{\epsfbox{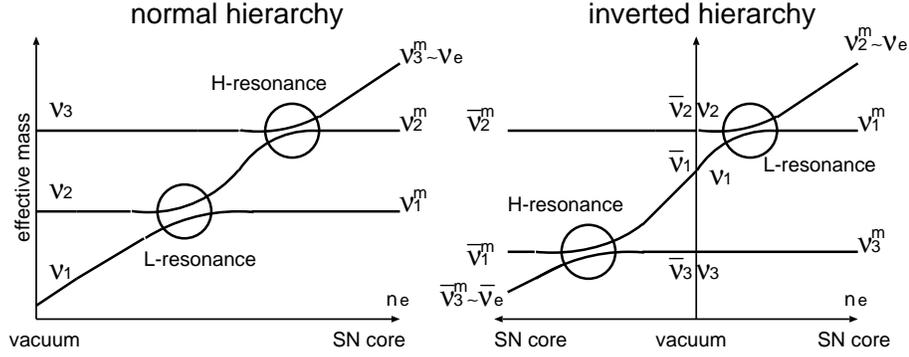}}
\caption{Schematic level crossing diagram for normal (left) and 
inverted (right) mass hierarchies. The circles shows the resonance points.}
\label{fig:level_crossing}
\end{figure}

We use a realistic model of a collapse-driven supernova
by the Lawrence Livermore group\cite{Wilson1986} to calculate
the neutrino luminosities and energy spectra, as we did in our previous
paper \cite{KT}.
The average energy of each flavor is:
\begin{equation}
\langle E_{\nu_{e}} \rangle \simeq 13 {\rm MeV}, \;\;
\langle E_{\bar{\nu}_{e}} \rangle \simeq 16 {\rm MeV}, \;\;
\langle E_{\nu_{x}} \rangle \simeq 23 {\rm MeV}.
\label{eq:nu_energy}
\end{equation}
Details of this original neutrino spectra are discussed by Totani et al.
\cite{Totani1998}
These neutrinos, which are produced in the high dense region of the iron
core, interact with matter before emerging from the supernova. Due to
the nonzero masses and the mixing in vacuum among various neutrino flavors,
flavor conversions can occur in supernova. When the mixing angle is small,
these conversions occur mainly in the resonance layer, where the density is
\begin{equation}
\rho_{\rm res} \simeq 1.4 \times 10^{6} {\rm g/cc}
\left( \frac{\Delta m^{2}}{1 {\rm eV}^{2}} \right)
\left( \frac{10 {\rm MeV}}{E_{\nu}} \right)
\left( \frac{0.5}{Y_{e}} \right)
\cos{2 \theta},
\end{equation}
where $\Delta m^{2}$ is the mass squared difference, $\theta$ is the
mixing angle, $E_{\nu}$ is the neutrino energy, and $Y_{e}$ is the mean 
number of electrons per baryon. Since the supernova core is dense enough, 
there are two resonance points in supernova envelope. One that occurs at 
higher density is called H-resonance and another is called L-resonance.
If the mass hierarchy is normal, both resonances occur in neutrino sector.
On the other hand, if the mass hierarchy is inverted, H-resonance occurs
in antineutrino sector and L-resonance occurs in neutrino sector.
The schematic level crossing diagram for normal and inverted mass hierarchies
are shown in Fig. \ref{fig:level_crossing}.

The dynamics of conversions including large mixing case is determined by the
adiabaticity parameter $\gamma$, which depend on the mixing angle and
the mass-squared difference between involved flavors:
\begin{equation}
\theta_{13} \; {\rm and} \; \Delta m^{2}_{13} \; {\rm at \; H-resonance},
\end{equation}
\begin{equation}
\theta_{12} \; {\rm and} \; \Delta m^{2}_{12} \; {\rm at \; L-resonance}.
\end{equation}
When $\gamma \gg 1$, the resonance is called 'adiabatic resonance' and 
the fluxes of the two involved mass eigenstate are completely exchanged. 
On the contrary, when $\gamma \ll 1$, the resonance is called
'nonadiabatic resonance' and the conversion does not occur. 
The dynamics of the resonance in supernova is studied in detail by
Dighe and Smirnov \cite{Dighe2000}.

\section{Method and Results}

In this section we describe the method of analysis and show the results.

\subsection{Conversion Probabilities}

In the framework of three-flavor neutrino oscillation, the time evolution
equation of the neutrino wave functions can be written as follows:
\begin{equation}
i\frac{d}{dt}\left(
	\begin{array}{ccc}\nu_e\\ \nu_{\mu}\\ \nu_{\tau}
	\end{array}\right)
= H(t)\left(
	\begin{array}{ccc}\nu_e\\ \nu_{\mu}\\ \nu_{\tau}
	\end{array}\right),
\end{equation}
\begin{equation}
H(t)\equiv 
U\left(
	\begin{array}{ccc}
		0 & 0 & 0\\
		0 & \Delta m^2_{12} /2E_{\nu} & 0\\
		0 & 0 & \Delta m^2_{13} /2E_{\nu}
	\end{array}\right)U^{-1} 
 	+\left(
	\begin{array}{ccc}
		A(t) & 0 & 0\\
		0 & 0 & 0\\
		0 & 0 & 0
	\end{array}\right),
\end{equation}
where $A(t)=\sqrt{2}G_{F}n_{e}(t)$, $G_{F}$ is Fermi constant, $n_{e}(t)$
is the electron number density, and $\Delta m^2_{ij}$ is the mass squared 
differences. In case of antineutrino, the sign of $A(t)$ changes.
Here U is the Cabibbo-Kobayashi-Maskawa (CKM) matrix:
\begin{equation}
U  =  \left(\begin{array}{ccc}
c_{12}c_{13} & s_{12}c_{13} & s_{13}\\
-s_{12}c_{23}-c_{12}s_{23}s_{13} & c_{12}c_{23}-s_{12}s_{23}s_{13} 
& s_{23}c_{13}\\
s_{12}s_{23}-c_{12}c_{23}s_{13} & -c_{12}s_{23}-s_{12}c_{23}s_{13} 
& c_{23}c_{13}
\end{array}\right)\label{mixing_matrix},
\end{equation}
where $s_{ij} = \sin{\theta_{ij}}, c_{ij} = \cos{\theta_{ij}}$ 
for $i,j=1,2,3 (i<j)$. We have here put the CP phase equal to zero in 
the CKM matrix.

By solving numerically these equations along the density profile of progenitor,
which is calculated by Woosley and Weaver\cite{Woosley1995},
we obtain conversion probabilities $P(\alpha \rightarrow \beta)$, i.e., 
probability that $\nu_{\alpha}$ at the center of the supernova becomes 
$\nu_{\beta}$ at the surface of the progenitor star.

In our previous paper\cite{KT}, we assumed the normal mass hierarchy and
took four models for neutrino oscillation parameters, the differences
being the solution of the solar neutrino problem (LMA or SMA) and
the magnitude of $\theta_{13}$. Here we take the following values:
\begin{eqnarray}
\sin^{2}{2 \theta_{12}} = 0.84, & \;\;\;\; & 
\Delta m^{2}_{12} = 7 \times 10^{-5} {\rm eV}^{2}, \nonumber \\
\sin^{2}{2 \theta_{23}} = 1.0, & \;\;\;\; & 
\Delta m^{2}_{23} = 3.2 \times 10^{-3} {\rm eV}^{2}.
\end{eqnarray}
Values of $\theta_{12}$ and $\Delta m^{2}_{12}$ are taken from
the global analysis of the solar neutrino observations and the KamLAND
experiment\cite{Bahcall2003} and correspond to the LMA solution
of the solar neutrino problem while those of $\theta_{23}$ and 
$\Delta m^{2}_{23}$ are taken from the analysis of the atmospheric
neutrino observation\cite{Fukuda1999}.
As to $\theta_{13}$, we take two fiducial values 
$\sin^{2}{2 \theta_{13}} = 0.043, \; 10^{-6}$ as we did in our previous
paper\cite{KT}. Later we will discuss the case of the other values.
Furthermore, we consider both normal and inverted hierarchy.
Consequently, there are four models and we call them normal-LMA-L,
normal-LMA-S, inverted-LMA-L and inverted-LMA-S.
The last character (L or S) represents the magnitude of $\theta_{13}$
(large or small). In our notation, 
$\Delta m^{2}_{ij} = m^{2}_{j} - m^{2}_{i}$ so that $\Delta m^{2}_{13} > 0$ 
in the normal hierarchy case and $\Delta m^{2}_{13} < 0$ in the inverted 
hierarchy case. Therefore, normal-LMA-L and inverted-LMA-L are different
only in the sign of $\Delta m^{2}_{13}$.

\begin{figure}[t]
\parbox{\halftext}{
\epsfxsize=6.5cm
\centerline{\epsfbox{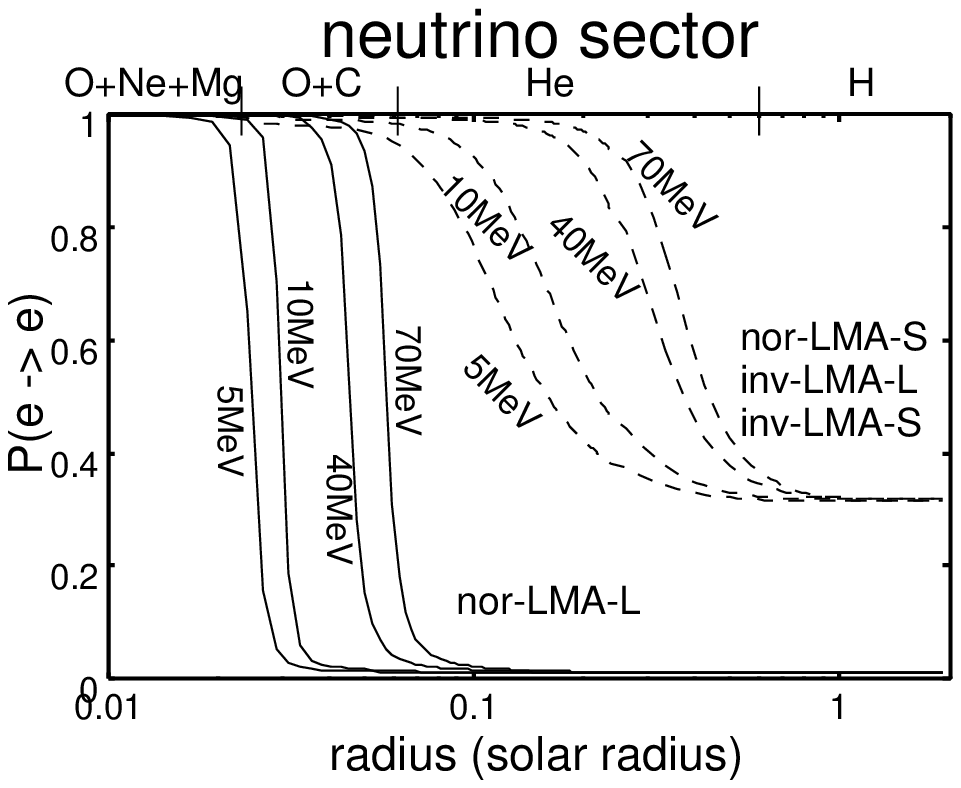}}}
\hspace{1cm}
\parbox{\halftext}{
\epsfxsize=6.5cm
\centerline{\epsfbox{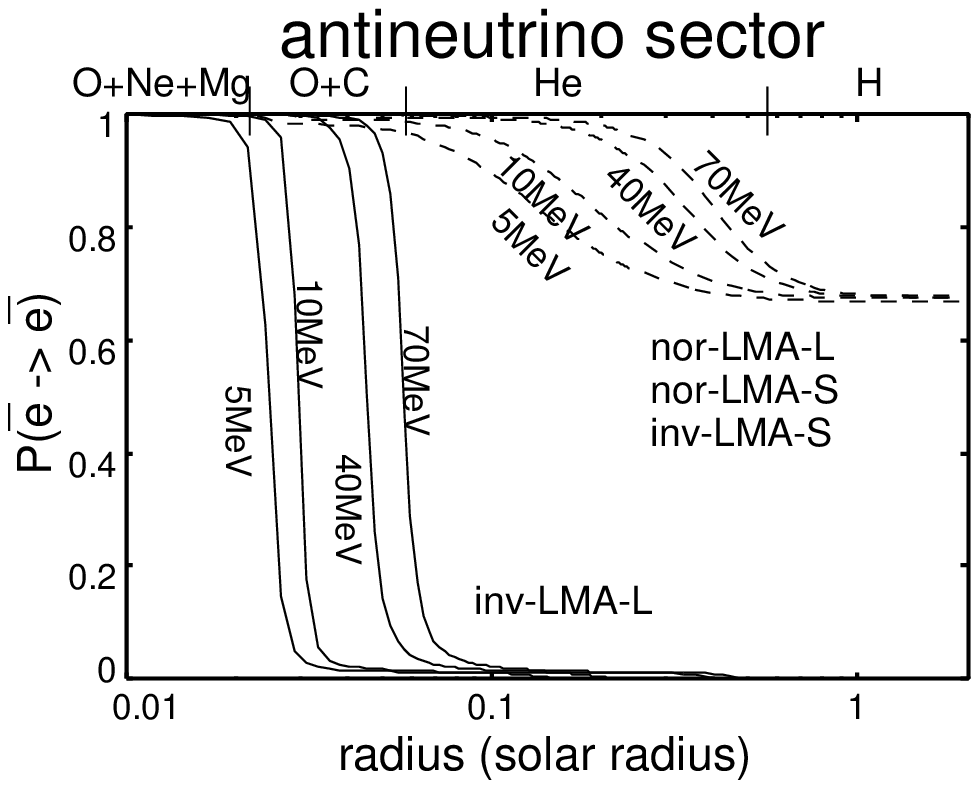}}}
\caption{Time evolution of conversion probabilities $P(e \rightarrow e)$ 
and $P(\bar{e} \rightarrow \bar{e})$.
Four lines of the same marking correspond to neutrino energy, 5MeV, 
10MeV, 40MeV, and 70MeV, respectively.}
\label{fig:conv_prob}
\end{figure}

We show in Fig. \ref{fig:conv_prob} demonstrations of conversion 
probabilities. The left figure is the 
time evolution of $P(e \rightarrow e)$, probability that $\nu_{e}$ remains
$\nu_{e}$, and the right figure is that of $P(\bar{e} \rightarrow \bar{e})$,
probability that $\bar{\nu}_{e}$ remains $\bar{\nu}_{e}$. 
Four curves for the same model correspond 
to the neutrino of energy, 5 MeV, 10 MeV, 40 MeV and 70 MeV, respectively.
It can be seen that H-resonance and L-resonance occur at the O+Ne+Mg
and He layer, respectively. In the neutrino sector, the conversion
probabilities for the inverted hierarchy are the same as those for
the model normal-LMA-S. This is because the H-resonance is completely
nonadiabatic when $\theta_{13}$ is very small, as in normal-LMA-S,
and this is phenomenologically as if the H-resonance is absent
as in the inverted hierarchy case. This logic also applies to why
normal-LMA-L, normal-LMA-S and inverted-LMA-S are degenerate in
the antineutrino sector, if we consider the H-resonance in
the antineutrino sector. These degeneracies are the origins of
the degeneracies that appear later in the event rates.

\subsection{Event Rates}

After obtaining the conversion probabilities, the neutrino fluxes at the Earth
are calculated by multiplying the conversion probabilities by the original
spectra and the distance factor $1/4 \pi d^{2}$. Here we take 10 kpc
for the distance $d$ between the Earth and the supernova.
Further, by multiplying these fluxes by the cross sections of the detection
interactions, the detector volume and the detector efficiency, we obtain
the event rates at the detectors. Here we consider two detectors: 
SuperKamiokande (SK) and SNO. Properties of these detectors and cross sections
used to calculate event rates are described in our previous paper \cite{KT}.
Unfortunate accident at SK lessened the detection efficiency at
low energy ($< 8$ MeV) but this cause negligible effect in the subsequent 
analysis.

\begin{figure}[t]
\parbox{\halftext}{
\epsfxsize=6.5cm
\centerline{\epsfbox{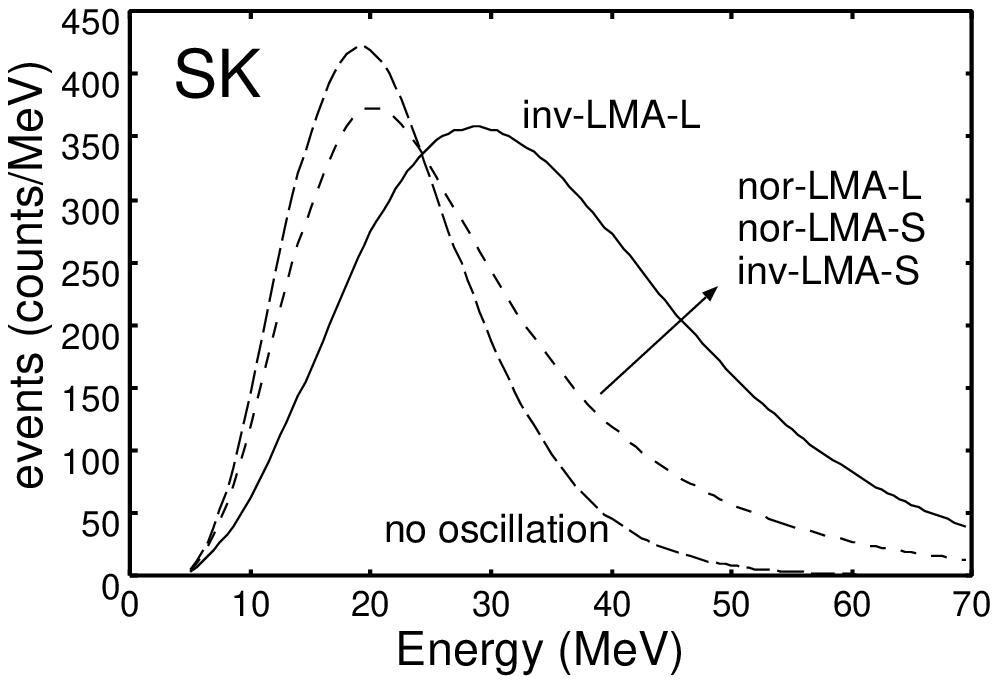}}}
\hspace{1cm}
\parbox{\halftext}{
\epsfxsize=6.5cm
\centerline{\epsfbox{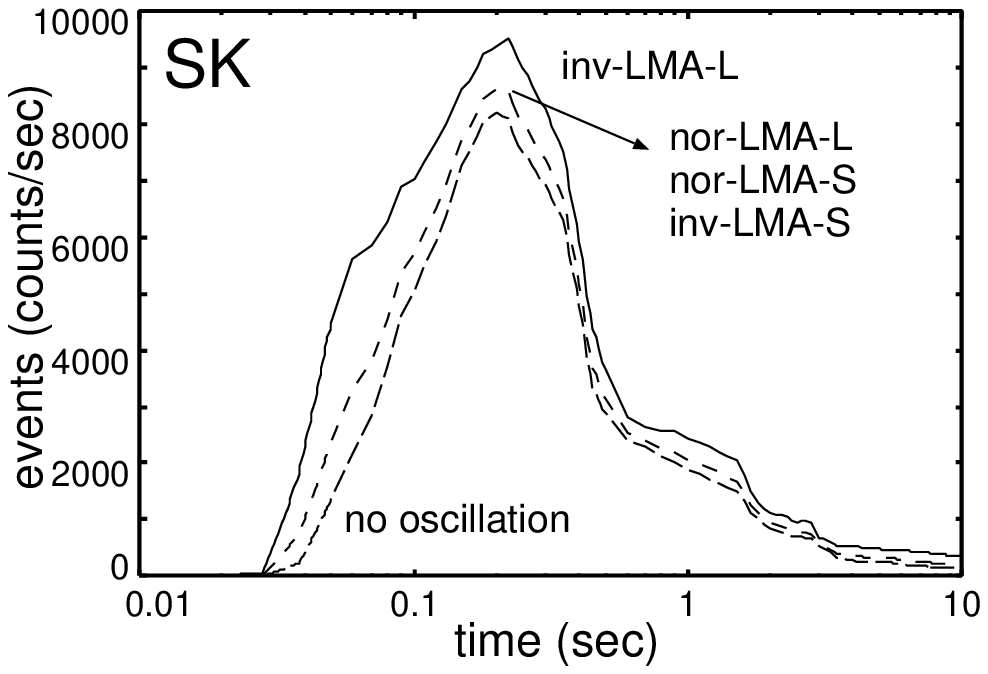}}}
\caption{The time-integrated energy spectra (left) and the time
evolution of the number of neutrino events (right) at SK. 
In these figures, only $\bar{\nu}_{e}p$ CC interaction is taken into account.}
\label{fig:SK}
\end{figure}

\begin{figure}[t]
\parbox{\halftext}{
\epsfxsize=6.5cm
\centerline{\epsfbox{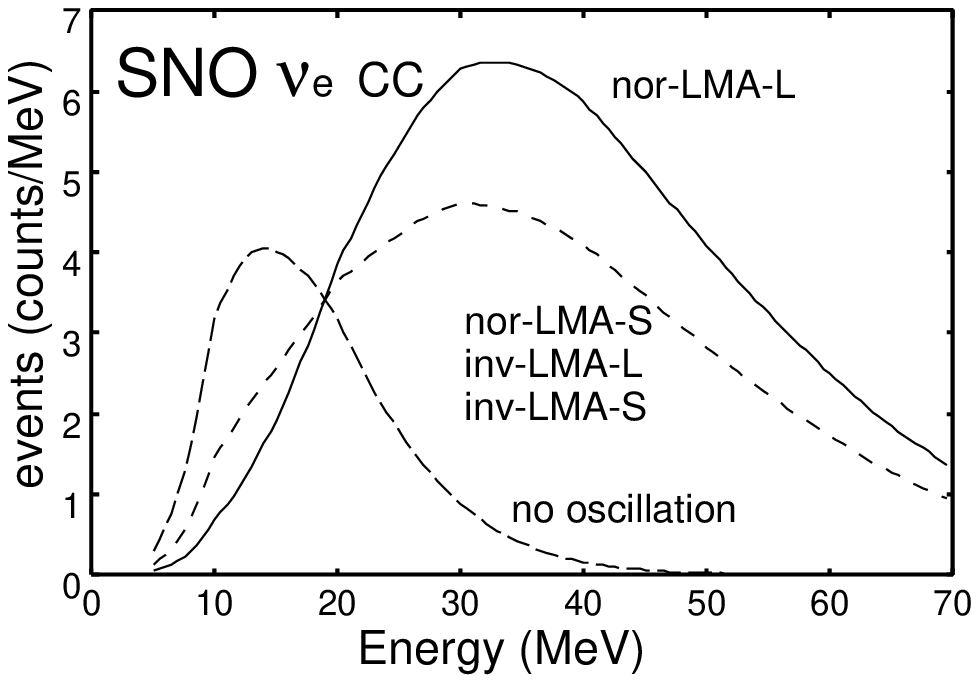}}}
\hspace{1cm}
\parbox{\halftext}{
\epsfxsize=6.5cm
\centerline{\epsfbox{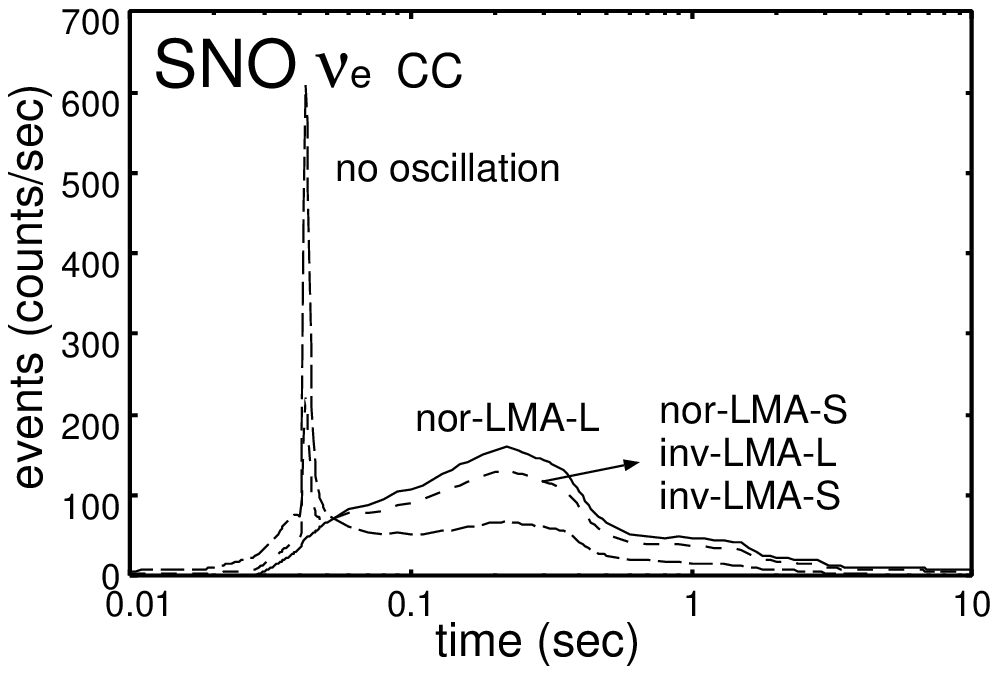}}}
\caption{The time-integrated energy spectra (left) and the time
evolution of the number of neutrino events (right) at SNO. In these figures,
only $\nu_{e}d$ CC interaction is taken into account.}
\label{fig:SNO}
\end{figure}

\begin{figure}[t]
\parbox{\halftext}{
\epsfxsize=6.5cm
\centerline{\epsfbox{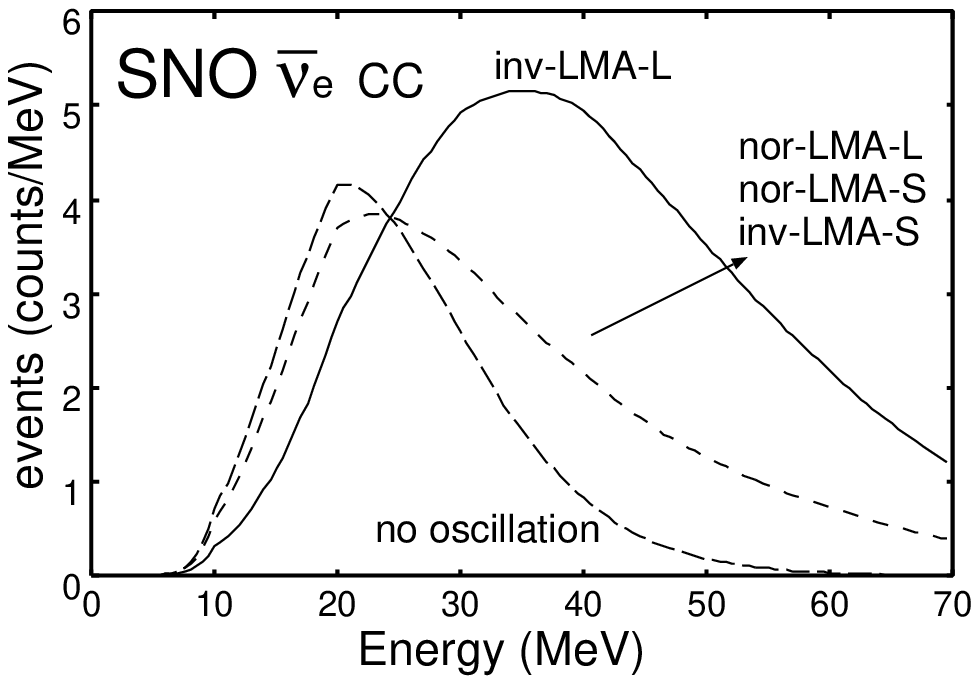}}}
\hspace{1cm}
\parbox{\halftext}{
\epsfxsize=6.5cm
\centerline{\epsfbox{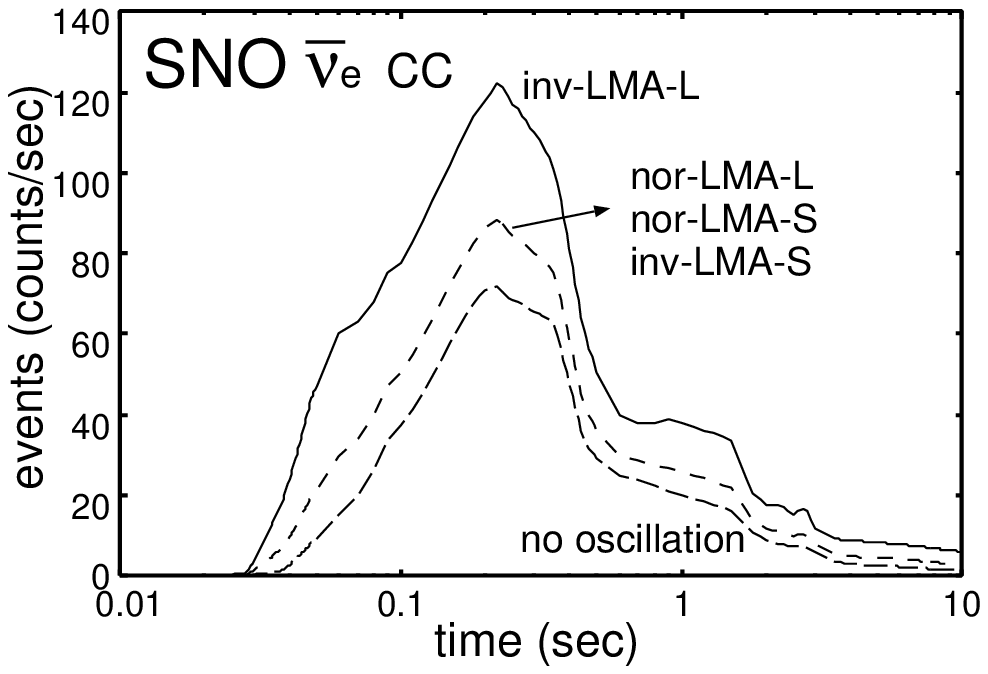}}}
\caption{The time-integrated energy spectra (left) and the time
evolution of the number of neutrino events (right) at SNO. In these figures,
only $\bar{\nu}_{e}d$ CC interaction is taken into account.}
\label{fig:SNO_anti}
\end{figure}

Fig. \ref{fig:SK} - \ref{fig:SNO_anti} show the time-integrated energy spectra 
(left) and the time evolution of the number of neutrino events (right) at 
SK and SNO ($\nu_{e}$ charged current (CC) events and 
$\bar{\nu}_{e}$ CC events), respectively. 
In Fig. \ref{fig:SK}, only $\bar{\nu}_{e}p$ CC interaction is taken 
into account. Event numbers of each interaction are shown 
in Table \ref{table:event_SK} and \ref{table:event_SNO}. In these tables, the 
contribution from neutronization burst phase is also shown. Here the 
neutronization burst phase means the period from 41msec to 48msec after 
the bounce.

\begin{table}
\caption{Number of events at SK \label{table:event_SK}}
\begin{center}
\begin{tabular}{c|cc|cc|c}
hierarchy & normal & & inverted & & \\
model & LMA-L & LMA-S & LMA-L & LMA-S  
& no osc\\ \hline
$\bar{\nu_e}p$ & 9459 & 9427 & 12269 & 9441 & 8036 \\
$\nu_ee^{-}$ & 186 & 171 & 171 & 171 & 132 \\
$\bar{\nu_e}e^{-}$& 46 & 46 & 56 & 46 & 42\\
$\nu_{\mu}e^{-}$ & 25 & 26 & 27 & 26 & 30 \\
$\bar{\nu_{\mu}}e^{-}$ & 24 & 23 & 12 & 23 & 24 \\
$\nu_{\tau}e^{-}$ & 25 & 26 & 26 & 26 & 30 \\
$\bar{\nu_{\tau}}e^{-}$& 24 & 23 & 12 & 23 & 24 \\
$O\nu_e$ & 297 & 214 & 297 & 214 & 31 \\
$O\bar{\nu_e}$ & 160 & 158 & 296 & 159 & 92 \\
\hline
total & 10245 & 10114 & 13084 & 10129 & 8441  \\ 
burst & 15.7 & 16.7 & 20.1 & 16.7 & 12.4
\end{tabular}
\end{center}
\end{table}

\begin{table}
\caption{Number of events (CC) at SNO
	\label{table:event_SNO}}
\begin{center}
\begin{tabular}{c|cc|cc|c}
hierarchy & normal & & inverted & & \\
model & LMA-L & LMA-S & LMA-L & LMA-S  
& no osc\\ \hline
$\nu_e d {\rm (CC)}$ & 237 & 185 & 185 & 185 & 68 \\ 
$\bar{\nu_e}d {\rm (CC)}$ & 118 & 117 & 190 & 117 & 82 \\
\hline 
total & 355 & 302 & 375 & 302 & 150 \\
burst & 0.6 & 1.1 & 1.1 & 1.1 & 2.1
\end{tabular}
\end{center}
\end{table}

\subsection{Distinction between Models\label{subsection:distinction}}

\begin{figure}[t]
\parbox{\halftext}{
\epsfxsize=6.5cm
\centerline{\epsfbox{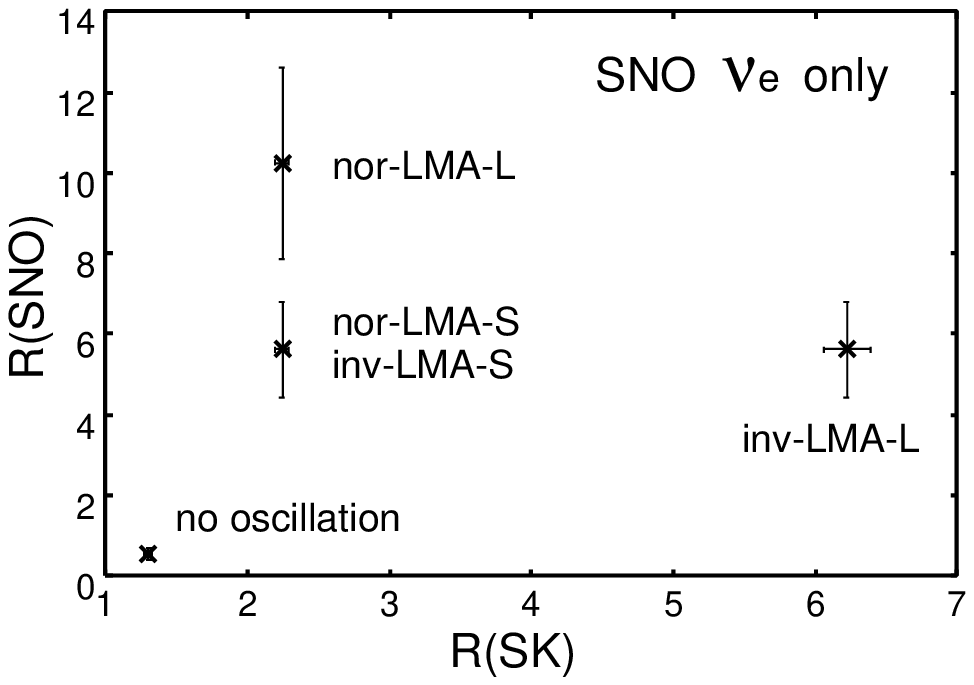}}}
\hspace{1cm}
\parbox{\halftext}{
\epsfxsize=6.5cm
\centerline{\epsfbox{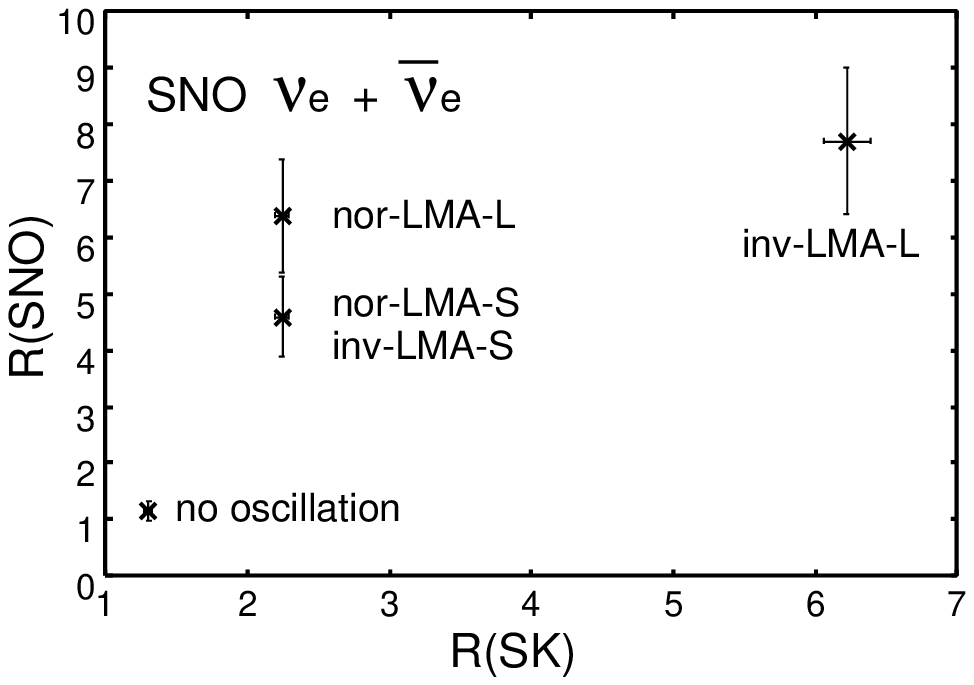}}}
\caption{Plots of $R_{\rm SK}$ vs $R_{\rm SNO}$.
In the left figure, only $\nu_{e} d$ CC events at SNO are considered for
$R_{\rm SNO}$. In the right figure we sum $\nu_{e} d$ CC events 
and $\bar{\nu}_{e} d$ CC events to calculate $R_{\rm SNO}$. The errorbars
show statistical errors only.
}
\label{fig:ratio}
\end{figure}

In general neutrino oscillation makes the $\nu_{e}$ and $\bar{\nu}_{e}$
spectra harder, since the original average energies of 
$\nu_{e}$ and $\bar{\nu}_{e}$ are smaller than that of $\nu_{x}$.
In other words neutrino oscillation produces high energy
$\nu_{e}$ and $\bar{\nu}_{e}$ from $\nu_{x}$. As a result, the high-energy
events increase and the low-energy events decrease. The boundary between
high energy and low energy is around 20 MeV. Note that how much
these increase and decrease are depends on the adiabaticity parametes, and then
the neutrino oscillation parameters, as can be seen in Fig. \ref{fig:SK},
\ref{fig:SNO} and \ref{fig:SNO_anti}. This feature can be used as a criterion
of the magnitude of the neutrino oscillation effects. We define
the following ratios of high-energy to low-energy events at both
detectors:
\begin{equation}
R_{\rm SK} \equiv \frac{\makebox{number of events at } 20<E<70\rm{MeV}}
{\makebox{number of events at } 5<E<20\rm{MeV}},
\end{equation}
\begin{equation}
R_{\rm SNO} \equiv \frac{\makebox{number of events at } 20<E<70\rm{MeV}}
{\makebox{number of events at } 5<E<20\rm{MeV}}.
\end{equation}
Note that the energy range for the definitions of $R_{\rm SK}$ and 
$R_{\rm SNO}$ are different from those in the previous paper \cite{KT}.
The plots of $R_{\rm SK}$ vs $R_{\rm SNO}$ are shown in Fig. \ref{fig:ratio}.
In the left figure, we consider only $\nu_{e} d$ CC events at SNO for
$R_{\rm SNO}$ assuming $\nu_{e} d$ CC event and $\bar{\nu}_{e} d$ CC event
can be distinguished completely. On the other hand, in the right figure
we assume that $\nu_{e} d$ CC event and $\bar{\nu}_{e} d$ CC event
can not be distinguished at all and we sum $\nu_{e} d$ CC events 
and $\bar{\nu}_{e} d$ CC events for $R_{\rm SNO}$. The error bars
represent the statistical errors.

Note that $\nu_{e}$ flux and $\bar{\nu}_{e}$ flux have essentially
different information about the neutrino oscillation parameters.
For example, inverted-LMA-L and inverted-LMA-S are distinguishable
from $\bar{\nu}_{e}$ events but are not from $\nu_{e}$ events.
So it is more effective to distinguish between models if $\nu_{e} d$
CC events and $\bar{\nu}_{e} d$ CC events at SNO can be distinguished 
perfectly. This can be clearly seen in Fig. \ref{fig:ratio}. 
In the left figure it is easier to distinguish between normal-LMA-L and 
(normal-LMA-S and inverted-LMA-S) than
in the right figure. But even in the left figure, it may be difficult
to distinguish between normal-LMA-L and (normal-LMA-S and
inverted-LMA-S) considering some ambiguities discussed in the following
sections.

\subsection{Uncertainties in the original spectra}

One of the crucial ingredients in this study is the flavor-dependences of
the original spectra. Smaller differences will make it more difficult
to distinguish between various models. There have been many numerical 
simulations of core collapse but the predicted luminosities and
temperatures of neutrinos are different from group by group.
Numerical model by the Livermore group\cite{Wilson1986}, 
adopted in this paper, has the great advantage that it covers the full 
evolution of supernova: from the core collapse over the explosion to 
the cooling phase of the protoneutron star, although it involves rather 
traditional treatments of neutrinos.

\begin{figure}[t]
\parbox{\halftext}{
\epsfxsize=6.5cm
\centerline{\epsfbox{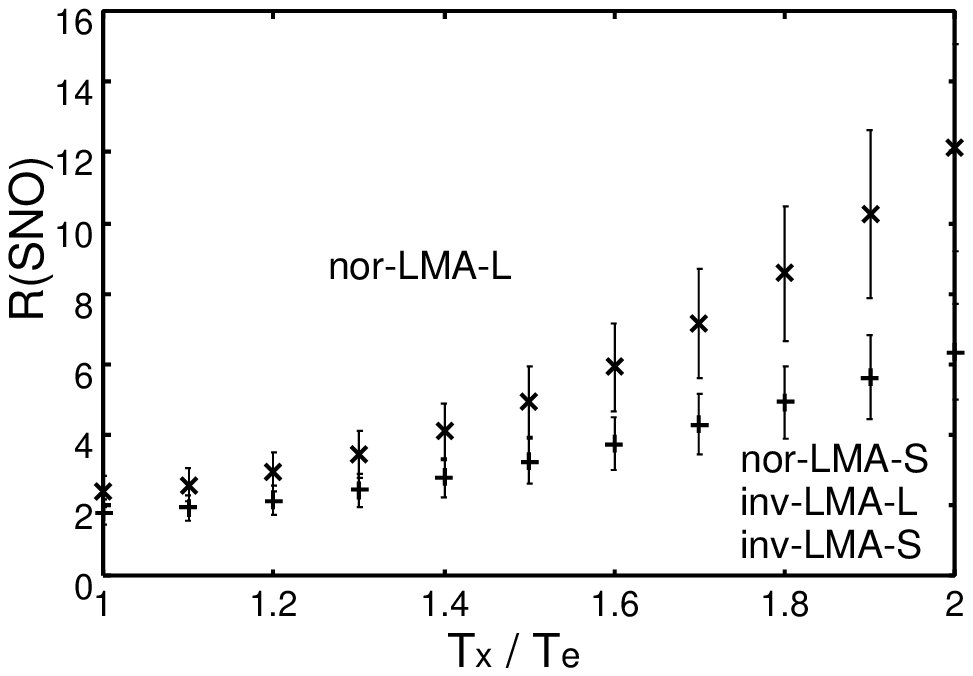}}}
\hspace{1cm}
\parbox{\halftext}{
\epsfxsize=6.5cm
\centerline{\epsfbox{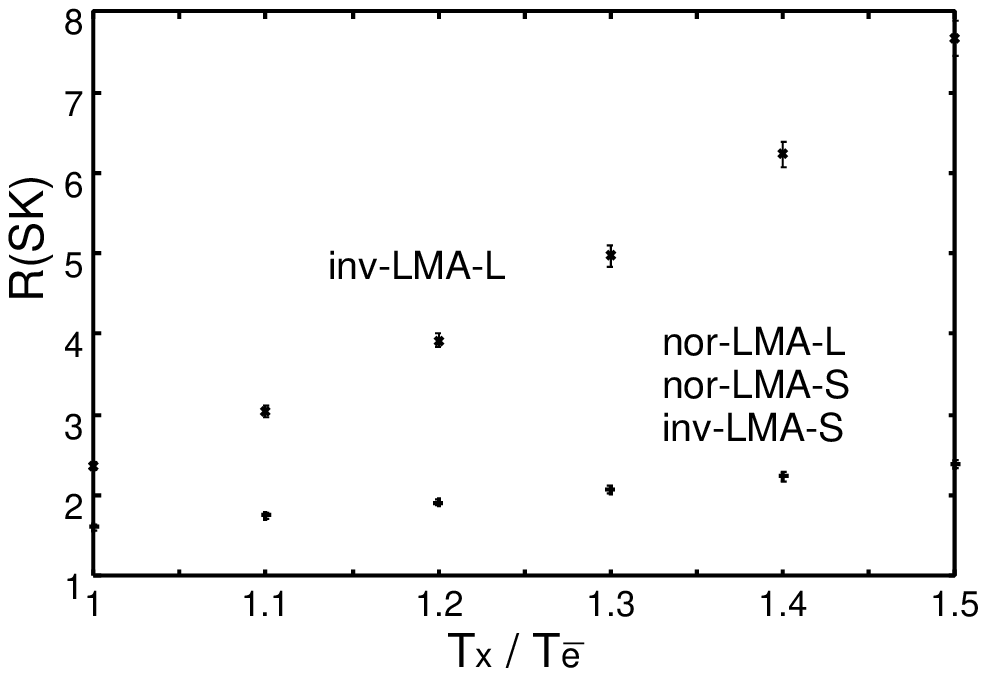}}}
\caption{Temperature ratio dependence of the event ratio at SNO and SK.
$T_{x}/T_{e}$ vs $R_{\rm SNO}$ in the left and 
$T_{x}/T_{\bar{e}}$ vs $R_{\rm SK}$ in the right.
In the left figure, only $\nu_{e}$ events are considered.
The errorbars show statistical errors only.}
\label{fig:spe_change}
\end{figure}

There are simulations with more sophisticated treatments of neutrinos
but they do not obtain explosions and neutrino fluxes for only less than
1 sec are available now\cite{Janka,Burrows}. 
These simulations predict less average-energy differences between flavors
compared to those of the Livermore group:
$E_{x}/E_{e} \sim 1.4$ \cite{Janka}, 1.7\cite{Burrows},
$E_{x}/E_{\bar{e}} \sim 1.1$ \cite{Janka}, 1.3\cite{Burrows},
where $E_{e}, E_{\bar{e}}$ and $E_{x}$ are the average energies of
$\nu_{e}, \bar{\nu}_{e}$ and $\nu_{x}$, respectively.
To see the effect of the uncertainties in the temperature differences
on our analysis, we perform similar analyses varying $\nu_{x}$ temperature 
as stated below and fixing $\nu_{e}$ and $\bar{\nu}_{e}$ temperatures.
This is relevant because various simulations agree well with respect to
$\nu_{e}$ and $\bar{\nu}_{e}$ temperatures.
The original spectra can be fitted by the ``pinched'' Fermi-Dirac
distribution,
\begin{equation}
f(E,T_{\alpha},\eta_{\alpha}) \propto 
\frac{E^{2}}{e^{E/T_{\alpha} - \eta_{\alpha}} + 1}, 
\end{equation}
where the temperature $T_{\alpha}$ and the pinching parameter $\eta_{\alpha}$
for each flavor ($\alpha = e, \bar{e}$ and $x$) are
\begin{eqnarray}
(T_{\alpha}, \eta_{\alpha}) 
& = & (2.8 {\rm MeV}, 2.6) \;\; {\rm for} \; \nu_{e}, \nonumber \\
&   & (4.0 {\rm MeV}, 2.5) \;\; {\rm for} \; \bar{\nu}_{e}, \nonumber \\
&   & (7.0 {\rm MeV}, 0) \;\; {\rm for} \; \nu_{x}.
\end{eqnarray}
We consider the following transformation of the distribution function of 
$\nu_{x}$ which has temperature $T_{x}$,
\begin{equation}
f(E,T_{x},\eta_{x}) \rightarrow \left(\frac{T_{x}}{T'_{x}}\right)^{2} 
f(E \frac{T'_{x}}{T_{x}},T'_{x},\eta_{x})
\label{eq:transformation}
\end{equation}
and we regard it as the distribution function of $\nu_{x}$ which
has temperature $T'_{x}$. Note that this transformation conserves
the total energy of $\nu_{x}$. Based on this spectrum with various
values of $T_{x}$, we obtain $R_{\rm SK}$ and $R_{\rm SNO}$
in the same way as in the previous subsections. Fig. \ref{fig:spe_change}
shows $T_{x}/T_{e}$ and $T_{x}/T_{\bar{e}}$ dependencies of 
$R_{\rm SNO}$ and $R_{\rm SK}$, respectively. In the left figure,
only $\nu_{e}$ events are considered to calculate $R_{\rm SNO}$. 
Predicted values by the Livermore group simulations are $T_{x}/T_{e} = 1.9$ 
and $T_{x}/T_{\bar{e}} = 1.4$, respectively. It should be noted that
$T_{x}/T_{e}$ and $T_{x}/T_{\bar{e}}$ are not independent parameters here
since we fix $T_{e}$ and $T_{\bar{e}}$. 

As can be seen, smaller values of these ratios result in smaller 
separation of event ratio between different models. In the left figure
the errorbars overlap each other if the temperature ratio is less than 1.5.
On the other hand, it seems that inverted-LMA-L and the others
can be discriminated independent of the temperature ratio.
The reason why $R_{\rm SK}$ is significantly different between
inverted-LMA-L and the others is that the pinching parameter is different
between $\nu_{e}$ and $\nu_{x}$. By the same reason, $R_{\rm SNO}$ for
normal-LMA-L is not the same those for the others even if $T_{x}/T_{e} = 1$.
Although the relevance of the transformation (\ref{eq:transformation}) is 
difficult to argue when $T_{x}/T_{\bar{e}}$ is nearly unity, the possibility 
of the discrimination between inverted-LMA-L and the others will be 
robust against the change of $\nu_{x}$ temperature of about $10 \%$.

\subsection{Dependence on $\theta_{13}$}

\begin{figure}[t]
\epsfxsize=6.5cm
\centerline{\epsfbox{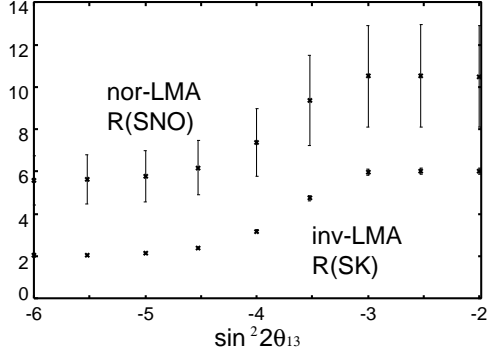}}
\caption{$\theta_{13}$
dependence of $R_{\rm SK}$ and $R_{\rm SNO}$. Only $\nu_{e}$ events
are taken into account to calculate $R_{\rm SNO}$.
The errorbars show statistical errors only.}
\label{fig:Ue3_change}
\end{figure}

So far we have considered only two extreme cases: H-resonance is
perfectly adiabatic and nonadiabatic. It is interesting to investigate
the intermediate cases. In Fig. \ref{fig:Ue3_change} we show $\theta_{13}$
dependence of $R_{\rm SK}$ and $R_{\rm SNO}$. Only $\nu_{e}$ events
are taken into account to calculate $R_{\rm SNO}$. Note that $R_{\rm SK}$
and $R_{\rm SNO}$ vary only in the case of inverted and normal
hierarchy, respectively, as will be expected from Fig. \ref{fig:ratio}.
In the case of normal hierarchy it will be difficult to determine
the value of $\theta_{13}$ due to large statistical errors but
will be possible to say whether it is very large or very small.
On the other hand, in the case of inverted hierarchy the overlap of 
the errorbars are small even in the intermediate cases.
If $\theta_{13}$ is rather large ($\sin^{2}{2 \theta_{13}} > 10^{-3}$) 
the mass hierarchy will be identified.

\subsection{Earth Effects\label{subsection:earth}}

Depending on the detector position, neutrino go through the Earth
and the matter effect inside the Earth can change the neutrino
spectra again. For the matter density of the Earth, neutrino oscillation
between only two light neutrinos is involved. Since the oscillation
length in the Earth, which depends on the neutrino energy, is the same order 
as the Earth radius for the neutrino parameter adapted here, the Earth
matter effect appear as a distosion in the spectra 
\cite{KTearth,KT3,Dighe2001,LunardiniSmirnov2001,DigheKeilRaffelt}.

\begin{figure}[t]
\parbox{\halftext}{
\epsfxsize=6.5cm
\centerline{\epsfbox{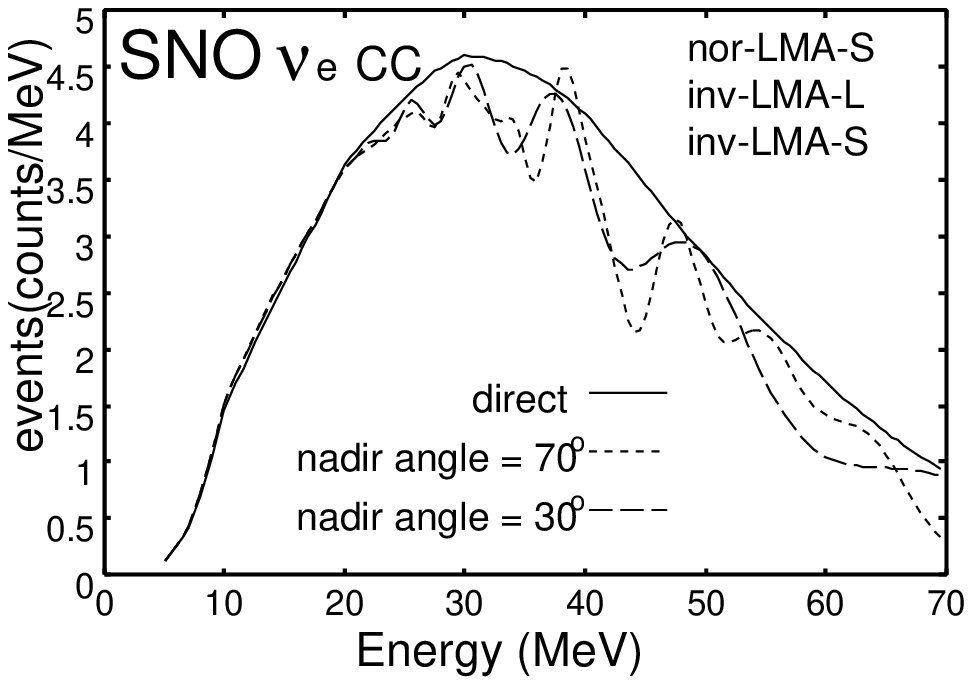}}}
\hspace{1cm}
\parbox{\halftext}{
\epsfxsize=6.5cm
\centerline{\epsfbox{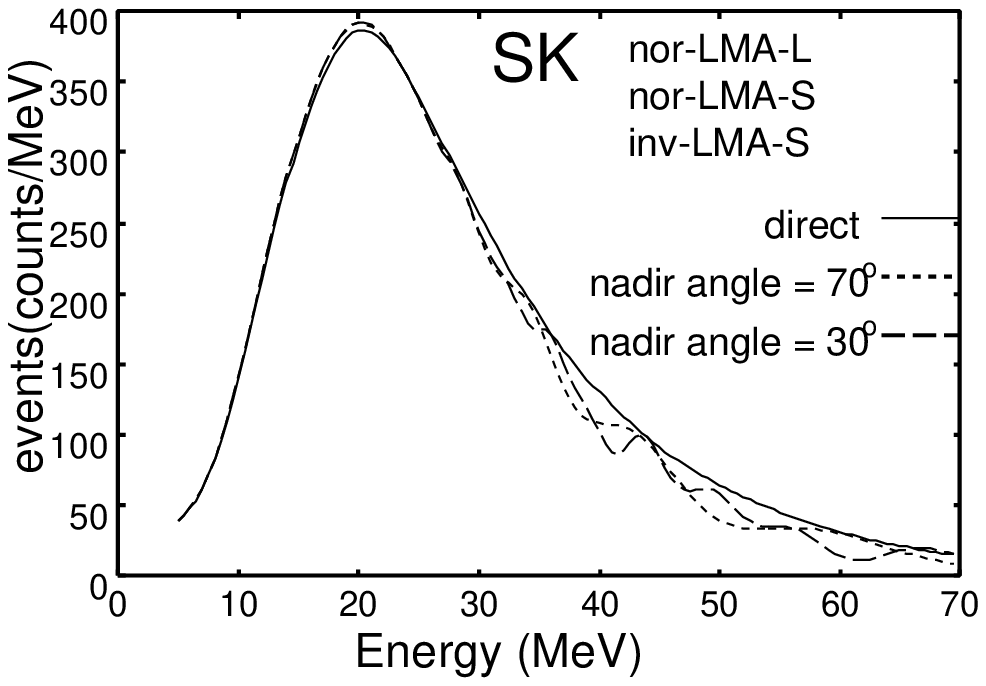}}}
\caption{Distorted spectra due to the Earth effects. The left figure 
is the $\nu_{e}$ spectra at SNO and the right figure is the 
$\bar{\nu}_{e}$ spectra at SK. Neutrino spectra in the case that
the nadir angle is $30^{\circ}$ and $70^{\circ}$ are shown with that
in absence of the Earth effects.}
\label{fig:Earth}
\end{figure}

The numerical process to calculate the Earth effect is the same as described 
above except that the density profile of the Earth is needed. We use a 
realistic density profile \cite{Dziewonski1981}. In Fig. \ref{fig:Earth} 
neutrino spectra in the case that the nadir angle is $30^{\circ}$ and 
$70^{\circ}$ are shown with that in absence of the Earth effects. The left 
figure is the $\nu_{e}$ spectra at SNO and the right figure is the 
$\bar{\nu}_{e}$ spectra at SK. The Earth effect is
absent in $\nu_{e}$ and $\bar{\nu}_{e}$ in case of normal-LMA-L
and inverted-LMA-L, respectively. This is because in these cases H-resonance
is perfectly adiabatic and low-energy neutrinos, which were originally
$\nu_{e}$ in the supernova core, are converted to the heaviest neutrino,
which is not involved in the matter oscillation in the Earth.
Thus the detection of the Earth effect will be helpful to distinguish
models, especially normal-LMA-L from normal-LMA-S and inverted-LMA-S.

The form of the distorted spectra depends on the nadir angle
of the neutrino path. The nadir angle $30^{\circ}$ and $70^{\circ}$ 
correspond to one of the path with which neutrino pass through core and mantle,
and only mantle, respectively. The detectability of the Earth effect
was discussed by us \cite{KT3} and it is shown that complementary
observation by SK, SNO and Large Volume Detector (LVD) is effective for 
its detection.

\section{Discussion and Conclusion\label{section:discussion}}

There are some ambiguities besides those descussed above.
The first is the direction of the supernova. If the supernova can
be observed optically, the direction can be known with enough accuracy.
But if the supernova is at the Galactic Center, it might be hidden by 
the large amount of gas and could not be seen optically. Pointing by the
electron scattering events of the supernova neutrino is studied
by several authors \cite{Beacom1999,Ando2001} and the accuracy is
expected to be $\sim 7^{\circ}$. More detailed analyses of the Earth 
effects have been studied considering the locations of the detectors
and the direction of the supernova \cite{KT3,LunardiniSmirnov2001}.

Another is the mass of the progenitor star. It affects the mass
of the iron core, which affects the neutrino spectra 
\cite{Burrows,MayleD,Mayle1987}. Study including the mass uncertainty
is now in progress but the preliminary result is that the mass
uncertainty is not important in our analysis \cite{KTmass}.

Recently effects of shock propagation on neutrino oscillation in supernova
have been studied 
\cite{SchiratoFuller2002,KTshock,LunardiniSmirnov2003,Fogli2003} 
and it was shown that some characteristic signatures may emerge as the 
shock propagates through the regions where matter-enhanced neutrino flavor 
conversion occurs. As we show \cite{KTshock}, shock propagation 
effect will be safely removed by taking only early-phase events into account.

We studied the effects of neutrino oscillation on supernova neutrino
in the case of the inverted mass hierarchy as well as the normal
mass hierarchy. Numerical analysis using a realistic supernova and 
presupernova model allowed us to discuss quantitatively a possibility to 
probe neutrino oscillation parameters. We showed that degeneracy exists 
only between normal-LMA-S and inverted-LMA-S if the Earth effect is taken 
into account and that $\theta_{13}$ can be well probed by SK if the neutrino 
mass hierarchy is inverted case. Errors due to the uncertainty of the original 
neutrino spectra are also estimated.

\section{Acknowledgments}

K.T.'s work is supported by Grant-in-Aid for JSPS Fellows.
K.S.'s work is supported by Grant-in-Aid for Scientific Research (S) No.
14102004 and Grant-in-Aid for Scientific Research on Priority Areas No.
14079202.

\end{document}